\documentclass[journal]{IEEEtran}
\usepackage{booktabs,siunitx}
\usepackage{tabularx,ragged2e}
\newcolumntype{L}{>{\RaggedRight\hangafter=1\hangindent=1em}X}
\usepackage{lipsum} 
\usepackage{xcolor,soul,framed} 
\usepackage{amssymb}
\colorlet{shadecolor}{yellow}
\usepackage{graphicx}
\usepackage{subcaption}
\graphicspath{{../pdf/}{../jpeg/}}
\DeclareGraphicsExtensions{.pdf,.jpeg,.png}

\usepackage[cmex10]{amsmath}
\usepackage{algorithm}
\usepackage{algpseudocode}
\usepackage{color}
\usepackage{mathtools}

\usepackage{array}
\usepackage{cite}
\usepackage{mdwmath}
\usepackage{mdwtab}
\usepackage{eqparbox}
\usepackage{url}
\usepackage{comment}
\usepackage{subcaption}
\usepackage{bbold}
\newcommand{\e}{{\mathrm e}}
\newcommand{\ji}{{\sf j}}

\begin{document}
\bstctlcite{IEEEexample:BSTcontrol}

\title{Off-grid Variational Bayesian Parameter Estimation for Fractional Delay-Doppler OTFS-ISAC}

\author{
    Van-Chung Luu, \IEEEmembership{Student Member, IEEE}
    and Duy H. N. Nguyen, \IEEEmembership{Senior Member, IEEE}%
    \thanks{This work has been submitted to the IEEE for possible publication. Copyright may be transferred without notice, after which this version may no longer be accessible.}
    \thanks{V.-C. Luu and D. H. N. Nguyen are with the Department of Electrical and Computer Engineering, San Diego State University, San Diego, CA 92182, USA. Emails: {\sffamily cluu1171@sdsu.edu} and {\sffamily duy.nguyen@sdsu.edu}.}%
}



\maketitle

\begin{abstract}
This letter proposes an off-grid variational Bayesian (OVB) method for fractional delay-Doppler (DD) estimation in OTFS-based integrated sensing and communication (ISAC) systems. To enable off-grid parameter estimation, the OTFS channel is reformulated using separable delay and Doppler steering vectors, and the corresponding phase variables are modeled by von Mises distributions. Closed-form variational updates provide posterior statistics for identifying significant paths and pruning redundant candidates, enabling automatic path-number estimation. Simulation results demonstrate that the proposed method achieves higher channel and parameter estimation accuracy than conventional fractional DD estimation approaches.


\end{abstract}


%
\IEEEpeerreviewmaketitle


\vspace{-0.5cm}
\section{Introduction}

Integrated sensing and communication (ISAC) has emerged as a key technology for future wireless networks by enabling communication and sensing to share spectrum, hardware, and waveform resources \cite{isac_overview}. This integration is particularly attractive for high-mobility and high-resolution applications such as vehicular networks, industrial automation, and 6G systems \cite{yuan2024otfs}. Among candidate waveforms, OTFS is well suited for ISAC because its sparse DD channel directly relates propagation parameters to range and velocity while providing strong robustness to high mobility through full time-frequency diversity \cite{OTFS_wave,OTFS_OFDM_comp}. Accordingly,  accurate DD-domain estimation of path gains, delays, and Doppler shifts is fundamental for both reliable data detection and sensing in OTFS-ISAC systems.

In high-mobility scenarios, the path delays and Doppler shifts are generally fractional with respect to the DD grid, causing energy leakage over neighboring DD bins and severe inter-path interference (IPI) \cite{CE_1,CE_2}. To address this issue, channel estimation for OTFS has been extensively investigated from both pilot-design and off-grid parameter-estimation perspectives. In \cite{CE_1}, an embedded pilot-aided scheme was proposed, where pilot, guard, and data symbols are jointly arranged in the DD domain to enable efficient channel estimation and data detection. To further handle fractional DD effects, \cite{CE_2} proposed an iterative path peak search method that refines off-grid path locations and mitigates IPI, thereby improving channel-estimation accuracy with manageable complexity. From a Bayesian learning perspective, off-grid sparse Bayesian learning (SBL) has been developed to jointly infer the on-grid support and off-grid delay-Doppler offsets, thereby exploiting DD-domain sparsity while reducing channel spreading effects \cite{CE_3}. Building on this probabilistic viewpoint, turbo inverse-free successive linear approximation VB inference (Turbo-IFSLA-VBI) further treats the sensing matrix as uncertain and jointly estimates the sparse channel, dynamic DD grid parameters, and support probabilities using successive linear approximation and turbo message passing \cite{CE_4}.

Beyond communication-oriented channel reconstruction, OTFS-ISAC requires the DD-domain paths to be estimated as physical parameters for sensing. In \cite{CE_8}, the delay-Doppler inter-path interference cancellation (DDIPIC) algorithm was proposed for joint communication-channel and radar-parameter estimation in fractional DD OTFS systems. DDIPIC models the received OTFS signal as a superposition of path-dependent atoms parameterized by path gain, delay, and Doppler shift, and estimates each path sequentially using a coarse integer-grid search followed by a fine fractional search. To reduce the complexity and improve the robustness of this sequential refinement strategy, several variants have been developed, including progressive residual-based IPI cancellation (P-IPIC) \cite{CE_5}, energy-leakage-guided sequential estimation \cite{CE_7}, and gradient-based off-grid refinement with interference cancellation \cite{CE_6}. 

Although these methods achieve accurate fractional DD parameter estimation, most existing OTFS-ISAC estimators mainly refine point estimates of fractional DD parameters through path search, cancellation, or local optimization, without fully exploiting posterior information for sparsity learning and model-order determination. This motivates a probabilistic inference framework for robust OTFS-ISAC parameter estimation. In this work, we develop a off-grid variational Bayesian framework for fractional OTFS-ISAC channel estimation. In addition, posterior-reliability-based pruning and duplicate-atom merging are incorporated to suppress inactive or redundant components, enabling model-order estimation without requiring prior knowledge of the true number of paths.


\section{System Model}

\begin{figure}[!th]
	\centering
\includegraphics[width=0.9\linewidth]{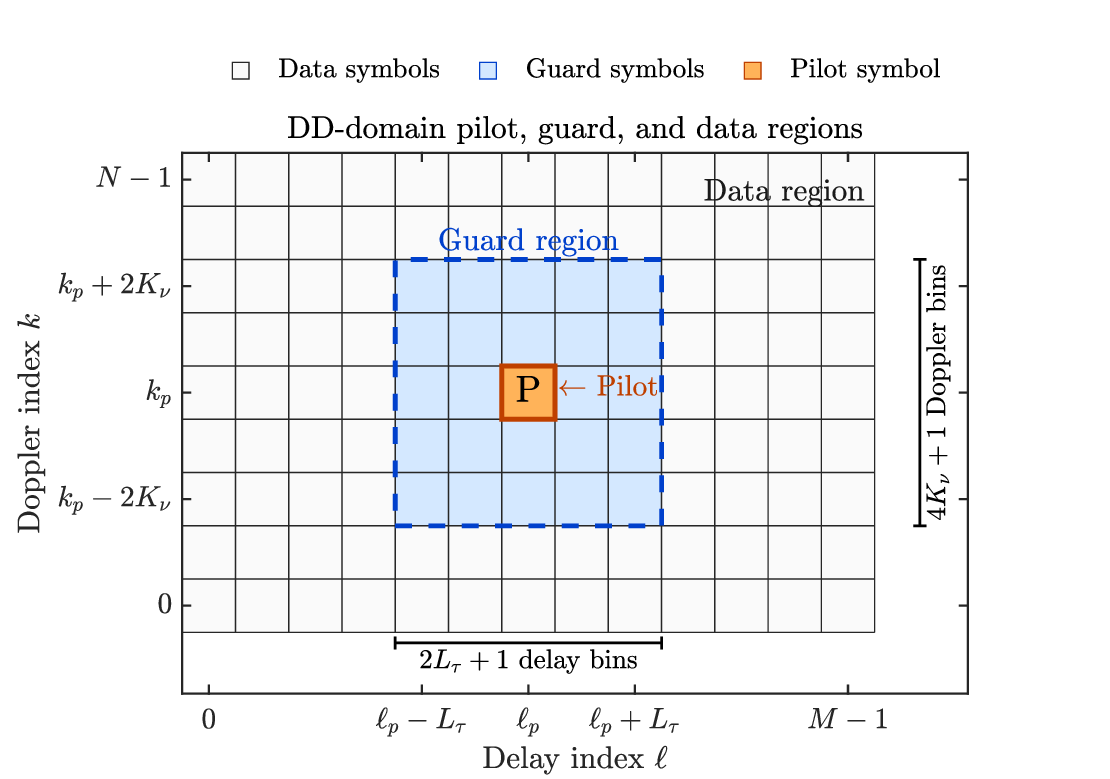}
	\caption{Illustration of OTFS symbol arrangement.}
	\label{fig: DD grid}
\end{figure}

We consider an OTFS frame with \(M\) subcarriers and \(N\) time slots. Equivalently, the DD grid contains \(M\) delay bins and \(N\) Doppler bins. Let \(\mathbf X_{\rm DD}\in\mathbb C^{M\times N}\) denote the transmitted DD-domain frame. The corresponding time-frequency (TF) frame is obtained through the unitary OTFS modulation transform
\begin{equation}
    \mathbf X_{\rm TF}
    =
    \mathbf F_M \mathbf X_{\rm DD}\mathbf F_N^{H},
    \label{eq:otfs_dd_to_tf}
\end{equation}
where \(\mathbf F_M\) and \(\mathbf F_N\) are unitary DFT matrices of sizes \(M\) and \(N\), respectively. The \((m,n)\)-th TF sample corresponds to the \(m\)-th subcarrier and the \(n\)-th OTFS time slot. An embedded sensing pilot is placed in the DD domain at \((\ell_p,k_p)=\left(\left\lfloor M/2\right\rfloor,\left\lfloor N/2\right\rfloor\right)\), with pilot amplitude \(x_p=\sqrt{MN}\). Following the standard embedded pilot–guard–data structure from \cite{CE_1}, a guard region is reserved around the pilot to isolate the pilot response from neighboring data symbols and thereby prevent pilot-data interference, as illustrated in Fig. \ref{fig: DD grid}. The receiver uses only the DD-domain samples inside the guarded observation set \(\Omega\). For monostatic ISAC sensing, the received echo is produced by \(P\) point targets or dominant scatterers. The \(p\)-th target is characterized by a complex reflection coefficient \(\alpha_p\), range \(R_p\), and radial velocity \(v_p\). Its physical round-trip delay and Doppler shift are
$\tau_p^{\rm phys}=2R_p/c$ and $f_{D,p}=2v_p f_c/c$, respectively, where
$c$ is the speed of light and $f_c$ is the carrier frequency. The corresponding
normalized fractional delay and Doppler bin indices are
$\tau_p=\tau_p^{\rm phys}/\Delta\tau=M\Delta f\,\tau_p^{\rm phys}$ and
$\nu_p=f_{D,p}/\Delta\nu=NT f_{D,p}$, where
$\Delta\tau=1/(M\Delta f)$ and $\Delta\nu=1/(NT)$. In practical propagation and radar sensing environments, the target delay and Doppler shifts are continuous-valued and generally do not align exactly with the discrete DD grid. Accurately modeling these off-grid components is important, since even small quantization errors in delay or Doppler can introduce range and velocity estimation errors, potentially leading to incorrect target localization. Therefore, the delay and Doppler shifts are decomposed as $\tau_p=\ell_{\tau,p}+\delta_{\tau,p}$ and $\nu_p=k_{\nu,p}+\delta_{\nu,p}$, where $\ell_{\tau,p}$ and $k_{\nu,p}$ denote the nearest integer delay and Doppler indices, while $\delta_{\tau,p}$ and $\delta_{\nu,p}$ represent the corresponding fractional off-grid offsets. Under the standard narrowband-per-subcarrier and blockwise constant target assumptions, the \(p\)-th target introduces a deterministic phase slope across the TF grid. Therefore, the TF-domain sensing channel at TF bin \((m,n)\) can be written as
\begin{equation}
    H_{\rm TF}[m,n]
    =
    \sum_{p=1}^{P}
    \alpha_p \,
    \e^{-\ji2\pi m\tau_p/M}
    \e^{\ji2\pi n\nu_p/N},
    \label{eq:tf_channel_scalar}
\end{equation}
for \(m=0,\ldots,M-1\) and \(n=0,\ldots,N-1\). Define the TF-domain delay and Doppler steering vectors as \(\mathbf a_{\tau}(\tau)=\left[1,\e^{-\ji2\pi\tau/M},\ldots,\e^{-\ji2\pi(M-1)\tau/M}\right]^T\) and \(\mathbf a_{\nu}(\nu)=\left[1,\e^{\ji2\pi\nu/N},\ldots,\e^{\ji2\pi(N-1)\nu/N}\right]^T\).
Then the vectorized TF-domain channel can be expressed as
\begin{equation}
    \mathbf h_{\rm TF}
    =
    \sum_{p=1}^{P}
    \alpha_p\,
    \mathbf b(\tau_p,\nu_p),
    \label{eq:tf_sparse_channel}
\end{equation}
where \(\mathbf b(\tau,\nu)=\mathbf a_{\nu}(\nu)\otimes\mathbf a_{\tau}(\tau)\).

The received TF-domain signal is modeled as the element-wise product between the transmitted TF frame and the TF-domain sensing channel,
\begin{equation}
    \mathbf y_{\rm TF}
    =
    \operatorname{diag}(\mathbf x_{\rm TF})
    \mathbf h_{\rm TF}
    +
    \mathbf w_{\rm TF},
    \label{eq:ytf_model}
\end{equation}
where \(\mathbf x_{\rm TF}=\operatorname{vec}(\mathbf X_{\rm TF})\in\mathbb C^{MN}\). After OTFS demodulation, the received DD-domain vector is
\begin{equation}
    \mathbf y_{\rm DD}
    =
    \mathbf T_{\rm rx}
    \operatorname{diag}(\mathbf x_{\rm TF})
    \mathbf h_{\rm TF}
    +
    \mathbf w_{\rm DD},
    \label{eq:ydd_model}
\end{equation}
where \(\mathbf T_{\rm rx}=\mathbf F_N^T\otimes\mathbf F_M^H\). Let \(\mathbf S_{\Omega}\) denote the row-selection matrix associated with the guarded DD observation set \(\Omega\). The guarded observation model is therefore
\begin{equation}
    \mathbf y_{\Omega}
    =
    \mathbf S_{\Omega}
    \mathbf T_{\rm rx}
    \operatorname{diag}(\mathbf x_{\rm TF})
    \mathbf h_{\rm TF}
    +
    \mathbf w_{\Omega}.
    \label{eq:guarded_observation_model}
\end{equation}
Substituting the sparse TF-domain sensing channel representation in \eqref{eq:tf_sparse_channel} into \eqref{eq:guarded_observation_model} gives
\begin{equation}
    \mathbf y_{\Omega}
    =
    \sum_{p=1}^{P}
    \alpha_p \,
    \boldsymbol\phi(\tau_p,\nu_p)
    +
    \mathbf w_{\Omega},
    \label{eq:guarded_sum_model}
\end{equation}
where the guarded OTFS sensing atom is defined as \(\boldsymbol\phi(\tau,\nu)=\mathbf S_{\Omega}\mathbf T_{\rm rx}\operatorname{diag}(\mathbf x_{\rm TF})\mathbf b(\tau,\nu)\). Since the number of targets is unknown, an overcomplete set of \(L\) candidate paths is introduced, where \(L\ge P\). Define \(\boldsymbol\Phi(\boldsymbol\tau,\boldsymbol\nu)=\left[\boldsymbol\phi(\tau_1,\nu_1),\ldots,\boldsymbol\phi(\tau_{L},\nu_{L})\right]\) and \(\boldsymbol\alpha=\left[\alpha_1,\ldots,\alpha_{L}\right]^T\).
The guarded OTFS-ISAC sensing model used for Bayesian inference is then
\begin{equation}
    \mathbf y_{\Omega}
    =
    \boldsymbol\Phi(\boldsymbol\tau,\boldsymbol\nu)
    \boldsymbol\alpha
    +
    \mathbf w_{\Omega},
    \label{eq:final_guarded_model}
\end{equation}
where \(\mathbf w_{\Omega}\sim\mathcal{CN}\left(\mathbf 0,N_0^{-1}\mathbf I\right)\), and \(N_0\) denotes the noise variance.
\vspace{-0.5cm}
\section{OVB OTFS-ISAC Parameter Estimation}

\subsection{Variational Bayesian Inference Framework}
Let
$\mathbf y$ denote the observed data and let
$\mathbf x=\{x_1,\ldots,x_m\}$ denote the latent variables. The goal is to
approximate the intractable posterior $p(\mathbf x|\mathbf y)$ by a simpler
distribution $q(\mathbf x)$. VB finds $q(\mathbf x)$ by minimizing
$\mathrm{KL}(q(\mathbf x)\,\|\,p(\mathbf x|\mathbf y))$, or equivalently by maximizing the evidence lower bound
$\mathrm{ELBO}(q)
=
\mathbb E_q[\ln p(\mathbf y,\mathbf x)]
-
\mathbb E_q[\ln q(\mathbf x)]$.
Under the mean-field approximation, the variational posterior is factorized as
$q(\mathbf x)=\prod_{i=1}^{m}q_i(x_i)$.
The optimal update is given as \cite{Bishop-2006}
\begin{equation}
    q_i^{\star}(x_i)
    \propto
    \exp
    \left\{
    \mathbb E_{q_{-i}}
    \left[
    \ln p(\mathbf y,\mathbf x)
    \right]
    \right\},
\end{equation}
where $q_{-i}=\prod_{j\neq i}q_j(x_j)$ denotes all variational factors except
$q_i(x_i)$.

\subsection{Hierarchical Bayesian Model}

We define the noise precision as $\gamma \triangleq 1/N_0$ and jointly infer it with the channel parameters from $\mathbf y_\Omega$. Using a mean-field VB approximation, the posterior is factorized as
\begin{align}
    p(\boldsymbol\alpha,
    \boldsymbol\vartheta_\tau,
    \boldsymbol\vartheta_\nu,
    \boldsymbol\beta,\gamma
    \!\mid\!\mathbf y_\Omega)
    &\approx
    q(\boldsymbol\alpha,
    \boldsymbol\tau,
    \boldsymbol\nu,
    \boldsymbol\beta,\gamma)
    \nonumber\\
    &=
    q(\gamma)
    \prod_{\ell=1}^{L}
    q(\alpha_\ell)
    q(\tau_{\ell})
    q(\nu_{\ell})
    q(\beta_\ell).
    \label{eq:mf_factorization}
\end{align}

Within the VB framework, the optimal variational densities in \eqref{eq:mf_factorization} are derived from the joint distribution $p(\mathbf y_\Omega,\boldsymbol\alpha,\boldsymbol\tau,\boldsymbol\nu,\boldsymbol\beta,\gamma)$, which factorizes as:

\begin{align}
    &p(\mathbf y_\Omega,\boldsymbol\alpha,
    \boldsymbol\tau,\boldsymbol\nu,
    \boldsymbol\beta,\gamma)
    \nonumber\\
    &\quad =
    p(\mathbf y_\Omega\!\mid\!\boldsymbol\alpha,
    \boldsymbol\tau,\boldsymbol\nu,\gamma)\,
    p(\boldsymbol\alpha\!\mid\!\boldsymbol\beta)\,
    p(\boldsymbol\tau)\,
    p(\boldsymbol\nu)\,
    p(\boldsymbol\beta)\,
    p(\gamma).
    \label{eq:joint_mfvb}
\end{align}
Here,
$p(\boldsymbol\alpha|\boldsymbol\beta)
=\prod_{\ell=1}^{L}p(\alpha_\ell|\beta_\ell)$
and
$p(\boldsymbol\beta)
=\prod_{\ell=1}^{L}p(\beta_\ell)$. To exploit the sparse nature of the OTFS channel, each path gain is assigned the hierarchical priors
$p(\alpha_\ell|\beta_\ell)=\mathcal{CN}(\alpha_\ell;0,\beta_\ell^{-1})$
and $p(\beta_\ell)=\Gamma(\beta_\ell;a_\beta,b_\beta)$,
where a large $\beta_\ell$ suppresses the $\ell$-th candidate path. The noise
precision follows
$p(\gamma)=\Gamma(\gamma;a_\gamma,b_\gamma)$. For given delay-Doppler parameters, the likelihood is written as
$p(\mathbf y_\Omega|\boldsymbol\alpha,\boldsymbol\tau,\boldsymbol\nu,\gamma)
=
\mathcal{CN}
\left(
\mathbf y_\Omega;
\boldsymbol\Phi(\boldsymbol\tau,\boldsymbol\nu)\boldsymbol\alpha,
\gamma^{-1}\mathbf I
\right)$,
where
$\boldsymbol\Phi(\boldsymbol\tau,\boldsymbol\nu)
=
[
\boldsymbol\phi(\tau_1,\nu_1),
\ldots,
\boldsymbol\phi(\tau_{L},\nu_{L})
]$.

\subsection{OVB Updates}
Each variational factor is obtained by taking the expectation of
$\ln p(\mathbf y_\Omega,\boldsymbol\alpha,
\boldsymbol\tau,\boldsymbol\nu,
\boldsymbol\beta,\gamma)$ with respect to all other variables. Let
$\boldsymbol\phi_\ell=\boldsymbol\phi(\tau_\ell,\nu_\ell)$. The residual for the $\ell$-th component is defined as
\begin{equation}
    \mathbf r_\ell
    =
    \mathbf y_\Omega
    -
    \sum_{i\neq \ell}
    \boldsymbol\phi_i\widehat{\alpha}_i .
    \label{eq:mfvb_residual_excluding}
\end{equation}

\emph{1) Update $\alpha_\ell$:}
By taking the expectation of \eqref{eq:joint_mfvb} over all variables except $\alpha_\ell$, the variational distribution of $\alpha_\ell$ is given by

\begin{align}
    q(\alpha_\ell)
    &\propto
    \exp\left\{
    \left\langle
    \ln p(\mathbf y_\Omega\!\mid\!
    \boldsymbol\alpha,
    \boldsymbol\tau,
    \boldsymbol\nu,
    \gamma)
    +
    \ln p(\alpha_\ell|\beta_\ell)
    \right\rangle_{-\alpha_\ell}
    \right\}
    \nonumber\\
    &\propto
    \mathcal{CN}
    \left(
    \alpha_\ell;
    \widehat{\alpha}_\ell,
    \sigma_{\alpha,\ell}^{2}
    \right).
\end{align}
The posterior variance and mean are respectively given by
$\sigma_{\alpha,\ell}^{2}
=
\left(
\widehat{\gamma}
\|\boldsymbol\phi_\ell\|_2^2
+
\widehat{\beta}_\ell
\right)^{-1}$
and
$\widehat{\alpha}_\ell
=
\sigma_{\alpha,\ell}^{2}\,
\widehat{\gamma}\,
\boldsymbol\phi_\ell^H
\mathbf r_\ell$.

\emph{2) Update $\tau_\ell$:}
Taking the expectation of the log-likelihood with respect to all latent variables
except $\tau_\ell$ yields
\begin{align}
    q(\tau_\ell)
    &\propto
    p(\tau_\ell)
    \exp
    \left\{
    \left\langle
    \ln p(\mathbf y_\Omega|
    \boldsymbol\alpha,\boldsymbol\tau,\boldsymbol\nu,\gamma)
    \right\rangle_{-\tau_\ell}
    \right\} \nonumber\\
    &\propto
    p(\tau_\ell)
    \exp
    \left(
    \Re\left[
    \boldsymbol\eta_{\tau,\ell}^H
    \mathbf a_{\tau}(\tau_\ell)
    \right]
    \right),
    \label{eq:q_tau_mfvb}
\end{align}
where
$\boldsymbol\eta_{\tau,\ell}
=
2\,\widehat{\gamma}\,\widehat{\alpha}_{\ell}^{*}\,
\mathbf B_{\tau,\ell}^{H}\mathbf r_{\ell}$
and
$\mathbf B_{\tau,\ell}
=
\mathbf S_{\Omega}\mathbf T_{\rm rx}
\operatorname{diag}(\mathbf x_{\rm TF})
\left(
\mathbf a_{\nu}(\widehat{\nu}_{\ell})
\otimes
\mathbf I_M
\right)$. To obtain a tractable circular approximation, we model $q(\vartheta_{\tau,\ell})$ as a von Mises (VM) distribution. Specifically, the posterior is
represented by a mean direction $\widehat{\tau_\ell}$ and a
concentration parameter $\kappa_{\tau,\ell}$ as,
\begin{equation}
    q(\tau_\ell)
    \approx
    \mathcal{VM}
    \left(
    \tau_\ell;
    \widehat{\tau}_\ell,
    \kappa_{\tau,\ell}
    \right).
\end{equation}
Therefore, the expected delay steering vector is
\begin{equation}
    \widehat{\mathbf a}_{\tau,\ell}
    =
    \mathbb E_{q(\tau_\ell)}
    \left[
    \mathbf a_{\tau}(\tau_\ell)
    \right]
    =
    \mathbf A_M(\kappa_{\tau,\ell})
    \mathbf a_{\tau}(\widehat{\tau}_\ell),
\end{equation}
where
$\mathbf A_M(\kappa)
=
\operatorname{diag}
\left(
1,
\frac{I_1(\kappa)}{I_0(\kappa)},
\ldots,
\frac{I_{M-1}(\kappa)}{I_0(\kappa)}
\right)$. For compact notation, let $\mathcal{H}{\tau}^{\mathrm{VM}}(\cdot)$ denote the Heuristic-2 procedure in \cite{badiu2017variational}, which approximates $q(\vartheta{\tau,\ell})$ by a single VM distribution and returns its mean direction and concentration parameter.

\emph{3) Update $\nu_\ell$:}
Similarly, the variational density of $\nu_\ell$ is
\begin{align}
    q(\nu_\ell)
    &\propto
    p(\nu_\ell)
    \exp
    \left\{
    \left\langle
    \ln p(\mathbf y_\Omega\!\mid\!
    \boldsymbol\alpha,\boldsymbol\tau,\boldsymbol\nu,\gamma)
    \right\rangle_{-\nu_\ell}
    \right\} \nonumber\\
    &\propto
    p(\nu_\ell)
    \exp
    \left(
    \Re\left[
    \boldsymbol\eta_{\nu,\ell}^H
    \mathbf a_{\nu}(\nu_\ell)
    \right]
    \right),
    \label{eq:q_nu_mfvb}
\end{align}
where
\begin{equation}
    \boldsymbol\eta_{\nu,\ell}
    =
    2\,\widehat{\gamma}\,\widehat{\alpha}_{\ell}^{*}\,
    \mathbf B_{\nu,\ell}^{H}\mathbf r_{\ell}.
    \label{eq:eta_nu_mfvb}
\end{equation}
The density $q(\nu_\ell)$ is approximated as
\begin{equation}
    q(\nu_\ell)
    \approx
    \mathcal{VM}
    (
    \vartheta_{\nu,\ell};
    \widehat{\nu}_\ell,
    \kappa_{\nu,\ell}
    ).
\end{equation}

\emph{3) Update $\beta_\ell$:}
The variational posterior of $\beta_\ell$ is given by
\begin{align}
    q(\beta_\ell)
    &\propto
    p(\beta_\ell)
    \exp\left\{
    \left\langle
    \ln p(\alpha_\ell|\beta_\ell)
    \right\rangle_{\alpha_\ell}
    \right\}
    \nonumber\\
    &\propto
    \Gamma\big(
    \beta_\ell;
    \widetilde a_{\beta,\ell},
    \widetilde b_{\beta,\ell}
    \big).
\end{align}
The posterior parameters are
\begin{equation}
    \widetilde a_{\beta,\ell}
    =
    a_\beta+1,
    \qquad
    \widetilde b_{\beta,\ell}
    =
    b_\beta
    +
    |\widehat{\alpha}_\ell|^2
    +
    \sigma_{\alpha,\ell}^{2}.
\end{equation}
Therefore, the posterior mean of $\beta_\ell$ is updated as
\begin{equation}
    \widehat{\beta}_\ell
    =
    \frac{a_\beta+1}
    {b_\beta
    +
    |\widehat{\alpha}_\ell|^2
    +
    \sigma_{\alpha,\ell}^{2}}.
    \label{eq:mfvb_beta_update}
\end{equation}

\emph{4) Update $\gamma$:}
The variational distribution of the noise precision is obtained as
\begin{align}
    q(\gamma)
    &\propto
    p(\gamma)
    \exp\left\{
    \left\langle
    \ln p(\mathbf y_\Omega|
    \boldsymbol\alpha,
    \boldsymbol\tau,
    \boldsymbol\nu,
    \gamma)
    \right\rangle_{-\gamma}
    \right\}
    \nonumber\\
    &\propto
    \Gamma(\gamma;a_\gamma,b_\gamma).
\end{align}
The posterior mean of the noise precision is updated as
\begin{equation}
    \widehat{\gamma}
    =
    \frac{a_\gamma+|\Omega|}
    {b_\gamma+\mathcal E},
    \label{eq:mfvb_gamma_update}
\end{equation}
where
$\mathcal E
=
\|\widehat{\mathbf r}\|_2^2
+
\sum_{\ell\in\mathcal A}
\sigma_{\alpha,\ell}^{2}
\|\boldsymbol\phi_\ell\|_2^2$,
and $\mathcal A$ denotes the current active set. 
\vspace{-0.4cm}
\subsection{Active-path Pruning and Duplicate Merging}
After each iteration, inactive candidate paths are pruned based on posterior energy and reliability. For $\mathcal L=\{1,\ldots,L\}$, define
\begin{equation}
    \widehat{E}_{\ell}
    \triangleq
    \mathbb{E}_{q}\!\left[|\alpha_{\ell}|^2\right]
    =
    |\widehat{\alpha}_{\ell}|^2+\sigma_{\alpha,\ell}^{2},
    \qquad
    \widehat{\chi}_{\ell}
    \triangleq
    \frac{|\widehat{\alpha}_{\ell}|^2}{\sigma_{\alpha,\ell}^{2}} .
    \label{eq:posterior_energy_reliability}
\end{equation}
The active set is updated as
\begin{equation}
    \mathcal A
    =
    \left\{
    \ell\in\mathcal L:
    \widehat{E}_{\ell}
    \geq
    \epsilon_E \max_{i\in\mathcal L}\widehat{E}_{i},
    \;
    \widehat{\chi}_{\ell}
    \geq
    \epsilon_{\chi} \max_{i\in\mathcal L}\widehat{\chi}_{i}
    \right\},
    \label{eq:active_set_update}
\end{equation}
where $0<\epsilon_E,\epsilon_{\chi}<1$. To avoid duplicate paths, highly correlated active atoms are merged. With $\widehat{\boldsymbol{\phi}}_{\ell}\triangleq\boldsymbol{\phi}(\widehat{\tau}_\ell,\widehat{\nu}_\ell)$, the normalized correlation is
\begin{equation}
    \rho_{i,\ell}
    =
    \frac{
    \left|
    \widehat{\boldsymbol{\phi}}_{i}^{H}
    \widehat{\boldsymbol{\phi}}_{\ell}
    \right|
    }{
    \|\widehat{\boldsymbol{\phi}}_{i}\|_2
    \|\widehat{\boldsymbol{\phi}}_{\ell}\|_2
    } .
    \label{eq:atom_correlation}
\end{equation}
If $\rho_{i,\ell}\geq\rho_{\rm th}$, the atom with smaller posterior energy is removed or absorbed into the stronger one.


\vspace{-0.4cm}
\subsection{Initialization}

The algorithm is initialized by a coarse greedy delay-Doppler search over
$\tau\in[0,L_{\tau}]$ and
$\nu\in[-K_{\nu},K_{\nu}]$. At each step, the atom maximizing the
normalized matched-filter score
$|\boldsymbol\phi(\tau,\nu)^H\mathbf r|^2/
\|\boldsymbol\phi(\tau,\nu)\|_2^2$
is selected, followed by a partial residual update. After selecting
$L_{\rm est}$ candidate paths, the gain vector is initialized by normalized
ridge least squares as
$\widehat{\boldsymbol\alpha}^{(0)}
=
(\widetilde{\boldsymbol\Phi}^H\widetilde{\boldsymbol\Phi}
+\lambda_r\mathbf I)^{-1}
\widetilde{\boldsymbol\Phi}^H\mathbf y_\Omega$,
where $\widetilde{\boldsymbol\Phi}$ denotes the column-normalized dictionary.
\vspace{-0.4cm}
\subsection{Computational Complexity of the OVB}
The initialization complexity is \(\mathcal{O}\!\left(LK_{\tau}K_{\nu}|\Omega|MN\right)\). In the OVB refinement stage, \(\bar{A}\le L\) denotes the average number of active paths, and \(K\) denotes the number of local grid points used in each one-dimensional delay or Doppler refinement. Thus, the refinement complexity is \(\mathcal{O}\!\left(\bar{A}K|\Omega|MN\right)\). Finally, the overall algorithm is summarized in Algorithm \ref{alg:mfvb_otfs_isac}.

\begin{algorithm}[t]
\caption{Proposed OVB Algorithm for Fractional OTFS-ISAC Parameter Estimation}
\label{alg:mfvb_otfs_isac}
\begin{algorithmic}[1]
\State \textbf{Input:} $\mathbf y_{\Omega}$, $\mathbf X_{\rm DD}$, $\Omega$,
$L$, and $T_{\max}$.
\State \textbf{Output:} $\widehat{\mathcal A}$, $\{\widehat{\alpha}_{\ell}, \widehat{\tau}_{\ell}, \widehat{\nu}_{\ell}\}_{\ell\in\widehat{\mathcal A}}$, and $\widehat{\mathbf h}_{\rm TF}$.
\State Initialize
$\{\widehat{\tau}_{\ell}^{1},
\widehat{\nu}_{\ell}^{1}\}_{\ell=1}^{L}$,
$\widehat{\boldsymbol\alpha}^{1}$,
$\widehat{\boldsymbol\beta}^{1}$,
$\widehat{\gamma}^{1}$, and
$\widehat{\mathcal A}^{1}=\{1,\ldots,L\}$.
\State Set
$\boldsymbol\phi_{\ell}^{1}
=
\boldsymbol\phi(
\widehat{\tau_\ell}^{1},
\widehat{\nu_\ell}^{1})$
and
$\boldsymbol\epsilon^{1}
=
\mathbf y_{\Omega}
-
\sum_{\ell\in\widehat{\mathcal A}^{1}}
\widehat{\alpha}_{\ell}^{1}
\boldsymbol\phi_{\ell}^{1}$.

\For{$t=1,2,\ldots,T_{\max}$}
    \For{$\ell\in\widehat{\mathcal A}^{t}$}
        \State $\mathbf r_{\ell}^{t}
        \leftarrow
        \boldsymbol\epsilon^{t}
        +
        \widehat{\alpha}_{\ell}^{t}
        \boldsymbol\phi_{\ell}^{t}$.


        \State $\boldsymbol\eta_{\tau,\ell}^{t}
        \leftarrow
        2\widehat{\gamma}^{t}
        (\widehat{\alpha}_{\ell}^{t})^{*}
        \mathbf B_{\tau,\ell}^{H}
        \mathbf r_{\ell}^{t}$.

        \State $[
        \widehat{\tau}_\ell^{t+1},
        \kappa_{\tau,\ell}^{t+1}]
        \leftarrow
        \mathcal{H}_{\tau}^{\mathrm{VM}}
        (\boldsymbol\eta_{\tau,\ell}^{t})$.

        \State $\boldsymbol\eta_{\nu,\ell}^{t}
        \leftarrow
        2\widehat{\gamma}^{t}
        (\widehat{\alpha}_{\ell}^{t})^{*}
        \mathbf B_{\nu,\ell}^{H}
        \mathbf r_{\ell}^{t}$.

        \State $[
        \widehat{\nu}_\ell^{t+1},
        \kappa_{\nu,\ell}^{t+1}]
        \leftarrow
        \mathcal{H}_{\nu}^{\mathrm{VM}}
        (\boldsymbol\eta_{\nu,\ell}^{t})$.

        \State $\boldsymbol\phi_{\ell}^{t+1}
        \leftarrow
        \boldsymbol\phi(
        \widehat{\tau}_\ell^{t+1},
        \widehat{\nu}_\ell^{t+1})$.

        \State $\sigma_{\alpha,\ell}^{2,t+1}
        \leftarrow
        \left(
        \widehat{\gamma}^{t}
        \|\boldsymbol\phi_{\ell}^{t+1}\|_2^2
        +
        \widehat{\beta}_{\ell}^{t}
        \right)^{-1}$.

        \State $\widehat{\alpha}_{\ell}^{t+1}
        \leftarrow
        \sigma_{\alpha,\ell}^{2,t+1}
        \widehat{\gamma}^{t}
        (\boldsymbol\phi_{\ell}^{t+1})^{H}
        \mathbf r_{\ell}^{t}$.

        \State $\boldsymbol\epsilon^{t}
        \leftarrow
        \mathbf r_{\ell}^{t}
        -
        \widehat{\alpha}_{\ell}^{t+1}
        \boldsymbol\phi_{\ell}^{t+1}$.
    \EndFor

    \State $\widehat{\beta}_{\ell}^{t+1}
    \leftarrow
    (a_{\beta}+1)/
    (b_{\beta}
    +
    |\widehat{\alpha}_{\ell}^{t+1}|^2
    +
    \sigma_{\alpha,\ell}^{2,t+1})$, $\forall \ell$.

    \State Update
    $\widehat{\mathcal A}^{t+1}$.

    \State $\boldsymbol\epsilon^{t+1}
    \leftarrow
    \mathbf y_{\Omega}
    -
    \sum_{\ell\in\widehat{\mathcal A}^{t+1}}
    \widehat{\alpha}_{\ell}^{t+1}
    \boldsymbol\phi_{\ell}^{t+1}$.

    \State $\mathcal E^{t+1}
    \leftarrow
    \|\boldsymbol\epsilon^{t+1}\|_2^2
    +
    \sum_{\ell\in\widehat{\mathcal A}^{t+1}}
    \sigma_{\alpha,\ell}^{2,t+1}
    \|\boldsymbol\phi_{\ell}^{t+1}\|_2^2$.

    \State $\widehat{\gamma}^{t+1}
    \leftarrow
    (a_{\gamma}+|\Omega|)
    /(b_{\gamma}+\mathcal E^{t+1})$.

    \State $\widehat{\mathbf h}_{\rm TF}^{t+1}
    \leftarrow
    \sum_{\ell\in\widehat{\mathcal A}^{t+1}}
    \widehat{\alpha}_{\ell}^{t+1}
    \mathbf b(
    \widehat{\tau}_\ell^{t+1},
    \widehat{\nu}_\ell^{t+1})$.
\EndFor

\end{algorithmic}
\end{algorithm}
\vspace{-0.4cm}
\section{Simulation Results}
This section presents the performance evaluation of the proposed OVB algorithm for fractional delay-Doppler parameter estimation in OTFS-ISAC systems. Reconstruction accuracy is evaluated using the normalized mean-squared error $\mathrm{NMSE}(\widehat{\mathbf z})=10\log_{10}\!\left(\|\mathbf z-\widehat{\mathbf z}\|_2^2/\|\mathbf z\|_2^2\right)$, where $\mathbf z$ denotes the true quantity of interest and $\widehat{\mathbf z}$ denotes its estimate. This metric is reported for the reconstructed channel $\widehat{\mathbf h}$ and for the estimated physical parameters $\widehat{\boldsymbol\alpha}$, $\widehat{\boldsymbol\tau}$, and $\widehat{\boldsymbol\nu}$, corresponding to the path gains, delays, and Doppler shifts, respectively. Unless otherwise specified, we consider an OTFS frame with $M=32$ delay bins and $N=16$ Doppler bins. The subcarrier spacing and carrier frequency are set to $\Delta f=15$ kHz and $f_c=5.1$ GHz, respectively. The maximum sensing range and velocity are $r_{\max}=2100$ m and $v_{\max}=100$ m/s, corresponding to $L_{\tau,\max}=7$ delay bins and $K_{\nu,\max}=4$ Doppler bins. The channel contains $P=4$ true paths, while the proposed algorithm initializes an overcomplete candidate list with $L=8$ paths. Parameters for pruning and merging \(\epsilon_E=2\times10^{-3}\), \(\epsilon_\chi=2\times10^{-3}\), \(\rho_{\rm th}=0.995\), \(\epsilon_\tau=0.25\), and \(\epsilon_\nu=0.25\). For each realization, the fractional path locations are generated by drawing \(r_p\sim\mathcal{U}(0,r_{\max})\) and \(v_p\sim\mathcal{U}(-v_{\max},v_{\max})\), and mapping them to DD-bin coordinates as \(\tau_p=(2r_p/c)/\tau_{\rm res}\) and \(\nu_p=(2v_pf_c/c)/\nu_{\rm res}\). The proposed OVB approach is compared with representative fractional DD channel-estimation methods, including off-grid SBL \cite{CE_3}, Turbo-IFSLA-VBI \cite{CE_4}, and P-IPIC \cite{CE_5}.

\begin{figure}[!t]
	\centering
\includegraphics[width=0.9\linewidth]{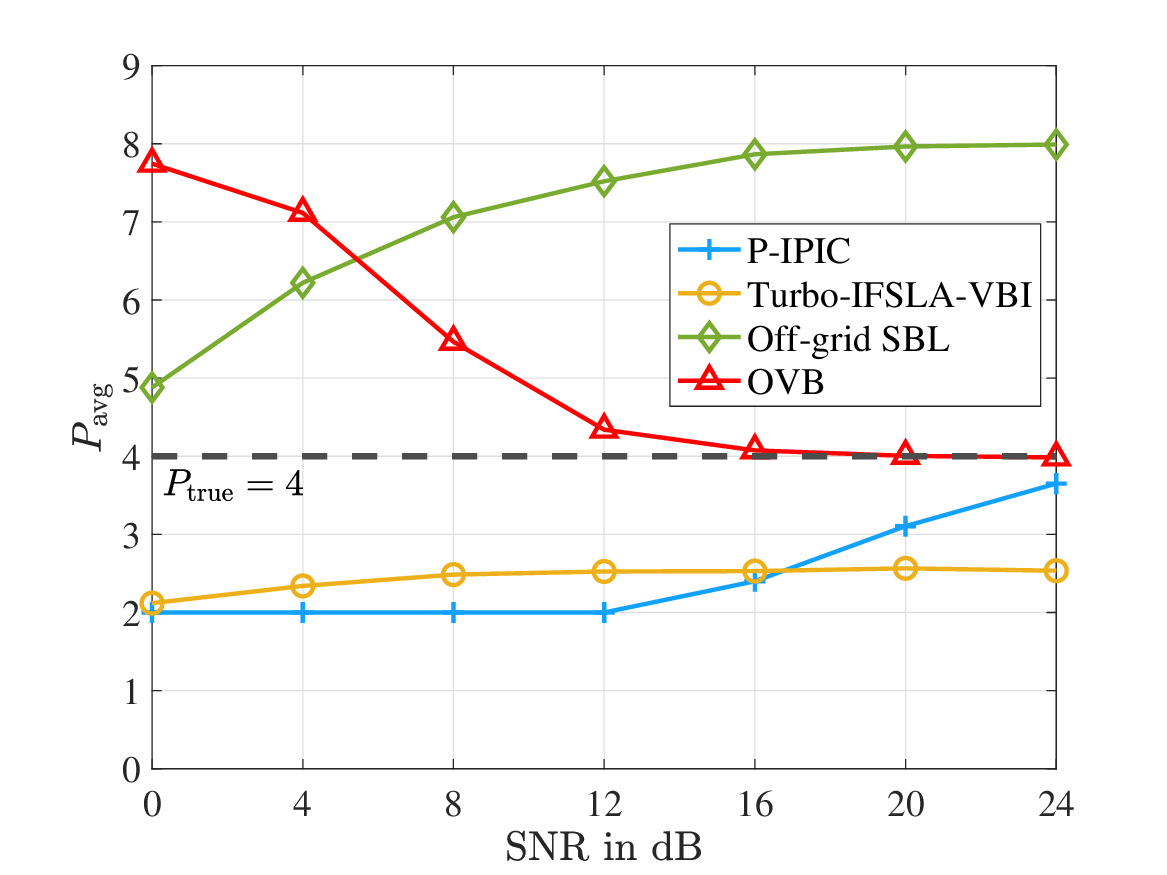}
	\caption{Average number of predicted paths.}
	\label{Predicted path}
\end{figure}

Fig.~\ref{Predicted path} shows the average number of predicted paths versus SNR. By exploiting the posterior statistics of the path gains, the proposed OVB method accurately estimates the model order, particularly in the high-SNR regime, where the predicted number of paths approaches the true value $P_{\rm true}=4$. In contrast, P-IPIC and Turbo-IFSLA-VBI tend to underestimate the number of paths, while off-grid SBL consistently overestimates it. These results demonstrate the advantage of using posterior uncertainty information for path-activity inference, rather than relying only on point estimates.

\begin{figure*}[t]
    \centering

    \begin{subfigure}[t]{0.32\textwidth}
        \centering
        \includegraphics[width=\linewidth]{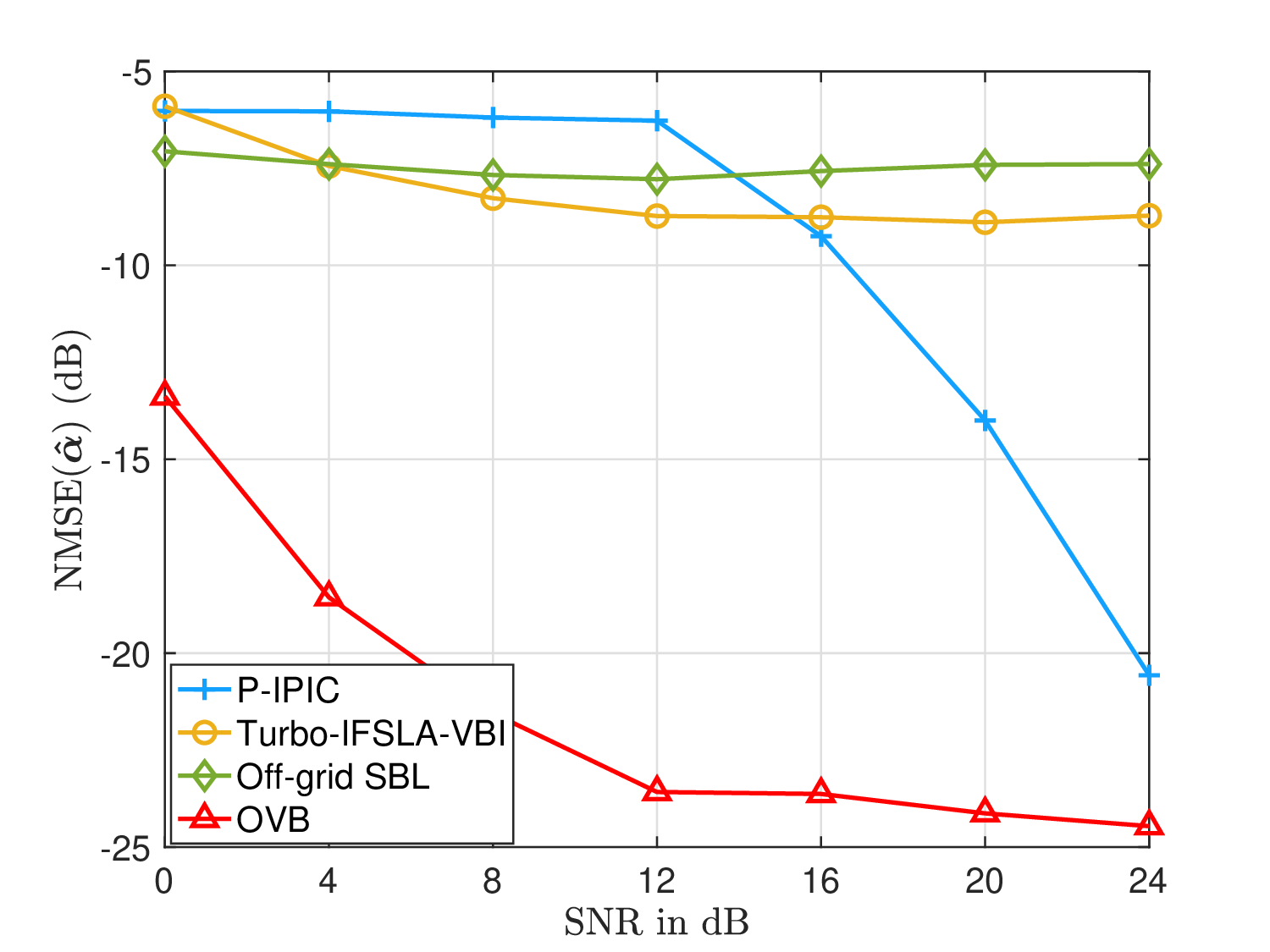}
        \caption{Channel gain}
        \label{fig:alpha}
    \end{subfigure}
    \hfill
    \begin{subfigure}[t]{0.32\textwidth}
        \centering
        \includegraphics[width=\linewidth]{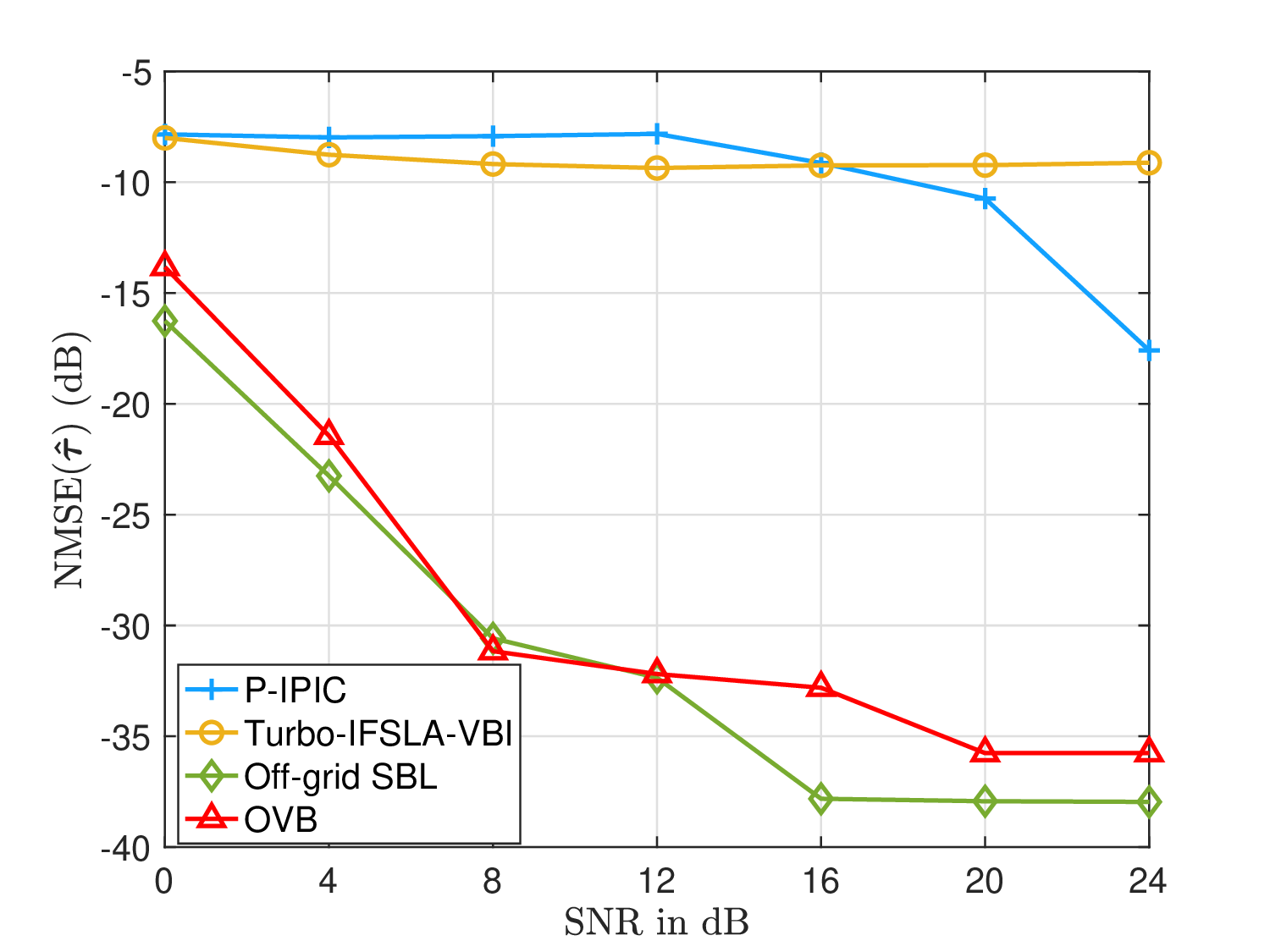}
        \caption{Delay}
        \label{fig:delay}
    \end{subfigure}
    \hfill
    \begin{subfigure}[t]{0.32\textwidth}
        \centering
        \includegraphics[width=\linewidth]{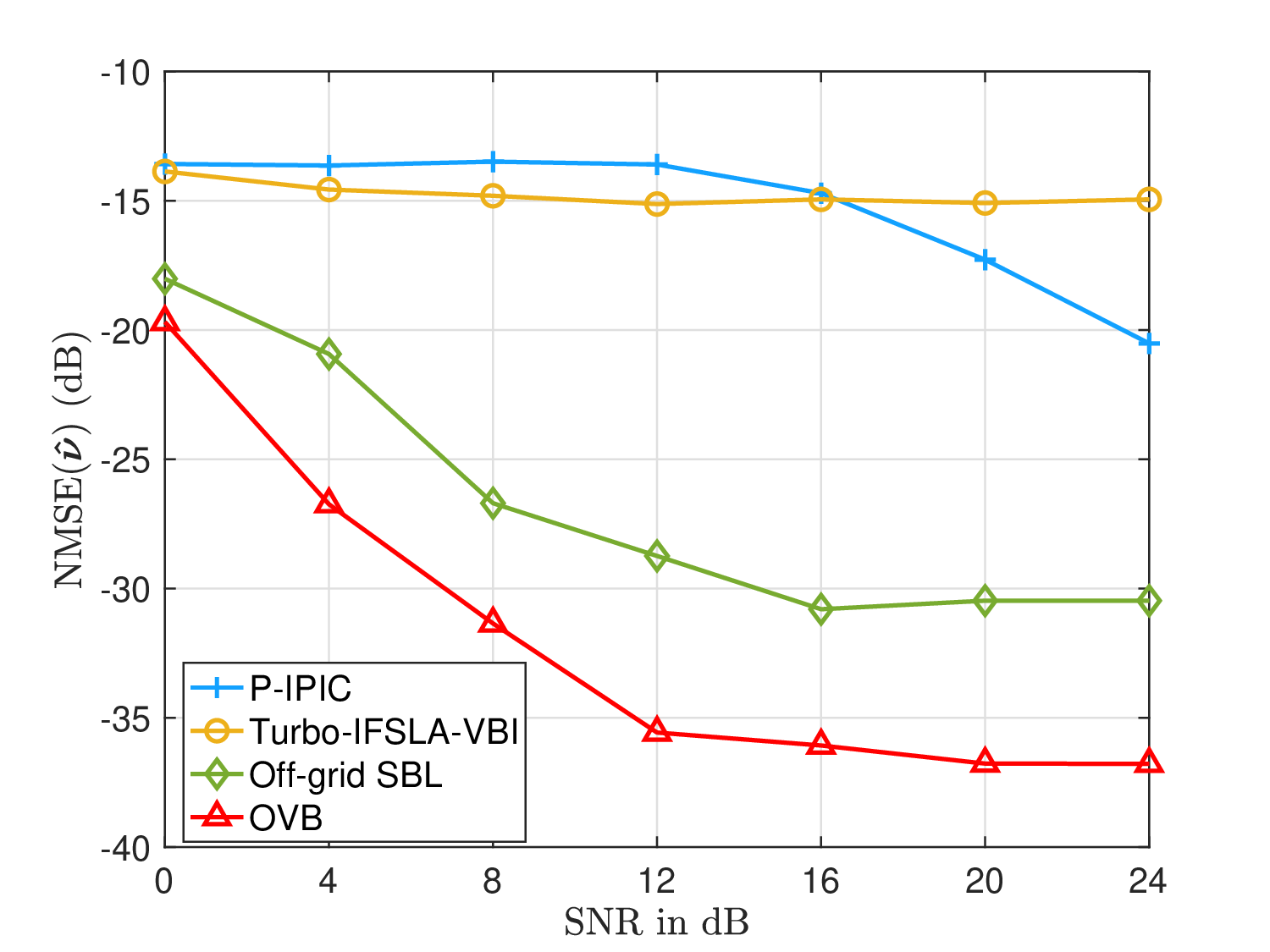}
        \caption{Doppler}
        \label{fig:doppler}
    \end{subfigure}

    \caption{NMSE of estimated channel gain, delay, and Doppler vs. SNR.}
    \label{fig:parameter_nmse}
\end{figure*}

Fig.~\ref{fig:parameter_nmse} compares the NMSE performance of the estimated complex gains $\boldsymbol\alpha$, delays $\boldsymbol\tau$, and Doppler shifts $\boldsymbol\nu$. The proposed OVB method achieves the best overall accuracy for all three parameters, with particularly large gains in complex-gain and Doppler estimation. This improvement is consistent with its ability to infer the model order accurately, which prevents both missing true paths and retaining redundant ones. Off-grid SBL provides competitive delay and Doppler estimates, but its tendency to overestimate the number of paths makes its performance less reliable, especially for gain estimation. P-IPIC improves at high SNR, whereas Turbo-IFSLA-VBI exhibits limited accuracy over the considered SNR range. These results show that accurate path-activity inference is critical for reliable fractional DD parameter estimation.

\begin{figure}[!th]
	\centering
\includegraphics[width=0.9\linewidth]{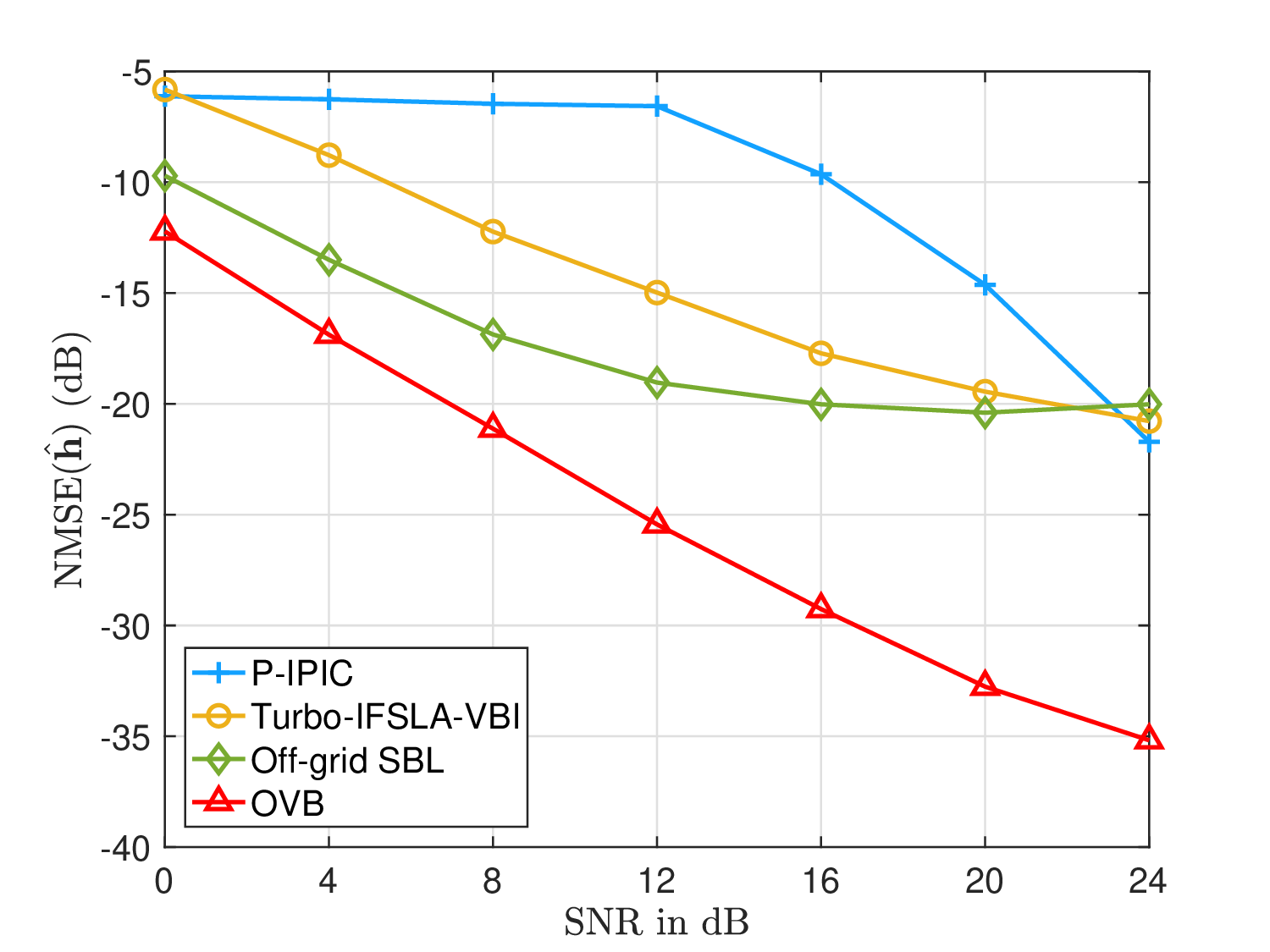}
	\caption{NMSE of channel reconstruction vs. SNR.}
	\label{fig: Channel estimation performance}
\end{figure}

Fig.~\ref{fig: Channel estimation performance} evaluates the channel reconstruction performance in terms of NMSE. The proposed OVB method achieves superior performance over all benchmark schemes across the considered SNR range, with increasingly significant gains at high SNRs. In contrast, the NMSEs of P-IPIC, Turbo-IFSLA-VBI, and off-grid SBL tend to converge at high SNR, indicating that their reconstruction accuracy is limited by residual model-order or parameter-estimation errors. These results further confirm the effectiveness of the proposed OVB method in achieving accurate and robust channel reconstruction for OTFS-ISAC systems.

\vspace{-0.4cm}
\section{Conclusion}

This letter has proposed an efficient OVB algorithm tailored for fractional DD channel estimation in OTFS-based ISAC systems. By modeling the off-grid delay and Doppler variables using von Mises distributions, the proposed method effectively handles fractional delay-Doppler effects in the DD domain. In addition, the derived variational update provide posterior statistics of the path gains, which enables identification of the active path set. Numerical simulations demonstrated that the proposed method outperforms existing fractional DD estimation schemes in terms of channel reconstruction and path-parameter estimation accuracy. These results highlight the potential of the OVB framework for robust and efficient OTFS-ISAC receiver design.

\bibliographystyle{IEEEtran}
\bibliography{IEEEabrv} 

\vfill


\end{document}